\documentclass[12pt]{article}
\usepackage{amssymb}
\usepackage{amsmath}
\usepackage{psfrag}
\usepackage{graphicx}

\newcommand{\cmt}[2]{\mbox{\begin{minipage}[t]{#1cm} #2 \end{minipage}}}
  
\newcommand{\Muserfunction}[1]{A}               %Mathematica-Importe

\newcommand{\PI}[3]{\big[\pi_{{#2}}(#1)\big]_{{#3}}}

\newcommand{\Sket}[3]{\big|~#1~#2~#3~\big>}

\newcommand{\barr}[1]{\begin{equation}\begin{array}{#1}}
\newcommand{\earr}{\end{array}\end{equation}}

\newcommand{\MC}[1]{\mathcal{#1}}

\newcommand{\eval}{\lambda_{\hat V}}

\def\ba{\begin{eqnarray}}
\def\ea{\end{eqnarray}}
\def\be{\begin{equation}}
\def\ee{\end{equation}}

\DeclareMathOperator{\sgn}{sgn}

\begin{document}

\begin{center}
{\normalsize \hfill Imperial/TP/2006/DR/02}\\

\vspace{40pt}
{ \Large \bf Spectral Analysis of the Volume Operator in Loop Quantum Gravity}
\footnote{Talk given by D. Rideout at the Eleventh Marcel Grossmann Meeting on General Relativity at the Freie U. Berlin, July 23 - 29, 2006.}

\vspace{20pt}

{\sl J. Brunnemann$^*$} and
{\sl D. Rideout$^\dagger$}

\vspace{15pt}
{\footnotesize

\begin{tabular}{cc}
\cmt{6}{
\begin{center}
$^*$Hamburg University\\
Mathematics Department\\
Bundesstrasse 55\\
20146 Hamburg, Germany\\[3mm]
% {\sl johannes.brunnemann@math.uni-hamburg.de}
\end{center}}
&
\cmt{6}{
\begin{center}
$^\dagger$Blackett Laboratory, \\
Imperial College,\\
London SW7 2AZ\\
United Kingdom.\\[3mm]
% {\sl d.rideout@imperial.ac.uk}
\end{center}}
\end{tabular}

\vspace{10pt}
Email: \\
{\sl $^*$johannes.brunnemann@math.uni-hamburg.de}, {\sl $^\dagger$d.rideout@imperial.ac.uk}
}
\vspace{10pt}

\begin{abstract}
We describe preliminary results of a detailed numerical analysis of the volume operator as formulated by Ashtekar and Lewandowski \cite{AL:QuantizedGeometry:VolumeOperators}. Due to    
a simplified explicit expression for its matrix elements\cite{VolumePaper}, it is possible for the first time to treat generic %(non-coplanar) 
vertices of valence greater than four. It is found that the vertex geometry characterizes the volume spectrum.      
\end{abstract}

\end{center}

\section{Introduction}\label{intro}

Loop Quantum Gravity\cite{TT:book} is an attempt to apply canonical quantization to General Relativity (GR). For this 
four dimensional spacetime $\MC{M}$ is foliated into an ensemble of three dimensional spatial hypersurfaces $\Sigma$.
GR can then be rewritten as an $SU(2)$ gauge theory with the canonical variables being densitized triads $E^a_i(x)$, and connections $A_b^j(y)$, encoding information on the induced metric $q$ on $\Sigma$. Here $(x,y)$ are points in $\Sigma$, $a,b=1,2,3$ are tensor indices, and $i,j=1,2,3$ are $SU(2)$-indices. 
In this treatment the theory is subject to constraints: three vector
% $C^a$ 
and one scalar constraint %$C$ 
ensuring invariance under diffeomorphisms within $\Sigma$ and deformations of $\Sigma$ within $\MC{M}$
respectively, and
three Gauss constraints $G_i$ which ensure invariance under $SU(2)$ gauge transformations.    
In the quantum theory one considers the integral of $A_b^j(y)$  over one dimensional edges $e\subset \Sigma_t$, that is the holonomies $h_e(A)=\int_e A$,  and fluxes $E_i(S)=\int_S \star E_i$ resulting from the integration of the dual of $E^a_i(x)$ over two dimensional surfaces $S \subset \Sigma_t$. 
Finite collections of edges are called a graph $\gamma$.  The edges mutually intersect at their beginning and end points, which are called the vertices $\{v\}|_\gamma$ of $\gamma$. 
The canonical pair $(h_e, E_i(S))$ can then be represented as multiplication and derivation operators respectively, on the space spanned by spin network functions (SNF)  $T_{\gamma\vec{j}\vec{m}\vec{n}}(h_{e_1}(A),\ldots,h_ {e_N}(A))=\prod_{e\subset \gamma}\PI{h_e}{j_e}{m_en_e}$ formulated with respect to  a particular $\gamma$. 
Each of the edges $(e_1,\ldots,e_N)$ of $\gamma$ carries a matrix element function $\PI{h_e}{j_e}{m_en_e}$ of an irreducible $SU(2)$-representation of weight $(j_1,\ldots,j_N)=:\vec{j}$ with matrix elements denoted by $(m_1,\ldots,m_N)=:\vec{m}$, $(n_1,\ldots,n_N)=:\vec{n}$. There is for each copy of $SU(2)$ attached to an edge $e\subset\gamma$ a one to one correspondence between the action  $\hat{E}_i(S) \PI{\cdot}{j}{mn}(\cdot)$ and the action of the usual angular momentum operator $J_i$ on an angular momentum state $\Sket{j}{m}{;n}$ with spin $(\sum_{i=1}^3 J_iJ_i)\Sket{j}{m}{;n}=j(j+1)\Sket{j}{m}{;n}$ and $J_3\Sket{j}{m}{;n}=m\Sket{j}{m}{;n}$, and an additional quantum  number $n$ which is not affected by the action of $J_i$.

\section{The Volume Operator}

As the theory is formulated classically in terms of the geometric objects $(A,E)$, it is possible to formulate a quantum version of the classical expression for the volume $V(R)$ of a spatial region $R\subset\Sigma$ given by %$V(R)=
$\int_R\sqrt{\det q}~d^3x=\int_R\sqrt{|\det E|}~d^3x$, where 
the classical identity $|\det E|=\det q$ is used (we assume $\det q>0$).
Upon quantization one obtains\cite{AL:QuantizedGeometry:VolumeOperators, VolumePaper}
$\hat{V}(R)T_{\gamma\vec{j}\vec{m}\vec{n}}(\cdot)
 =\ell_P^3 \sum_{\{v\}|_{\gamma\cap R}} \sqrt{\big|Z\sum_{IJK}\epsilon(IJK)\hat{q}_{IJK}\big|}
 T_{\gamma\vec{j}\vec{m}\vec{n}}(\cdot)
 $. Here $\ell_P$ is the Planck length, $Z$ is a constant and $\hat{q}_{IJK}:=4\epsilon^{ijk}J^{e_I}_i J^{e_J}_j J^{e_K}_k$ is a polynomial of operators,  $J^{e_I}_i$ denoting the $i$-component of angular momentum acting on the $SU(2)$-copy attached to the edge $e_I$. 
In the action of $\hat{V}(R)$
the classical integration $\int_R$ is replaced by a sum $\sum_{\{v\}|_{\gamma\cap R}}$ over vertices $v$ of $\gamma$ contained in $R$, so volume is concentrated at vertices only. At each vertex $v$ of $\gamma$ one obtains a matrix $\hat{q}_{IJK}$ for each triple $e_I\cap e_J \cap e_K=v$ of edges incident at $v$. These matrices are added with prefactors $\epsilon(IJK):=\sgn{(\det{({\dot{e}}_I(v),{\dot{e}}_J(v),{\dot{e}}_K(v)}))}=0,\pm 1$, which carry spatial diffeomorphism invariant information on the orientation of the triple of edge tangent vectors ${\dot{e}}_L(v):=\frac{d}{ds}e_L(s)|_v$ for each edge $e_L$ at $v$, with curve parameter $s$. If the tangents are coplanar then $\epsilon(IJK)=0$.
Taking the matrix sum we obtain a purely imaginary antisymmetric matrix
with real eigenvalues $\lambda_q$ (which come in pairs $\pm |\lambda_q|$ or are 0) and eigenstates $T_{\lambda_q}$ (linear combinations of the $T_{\gamma\vec{j}\vec{m}\vec{n}}(\cdot)$). $\hat{V}(R)$ then has $T_{\lambda_{\hat{V}}}=T_{\lambda_q}$ as eigenstates with according eigenvalues $\lambda_{\hat{V}}=\sqrt{|\lambda_q|}$.

\section{Spectral Analysis}

\begin{figure}[tbhp]
\begin{center}
  \psfrag{number}{$N_{\eval}$}
  \psfrag{eval}{$\eval$}
\includegraphics[width=5.5in]{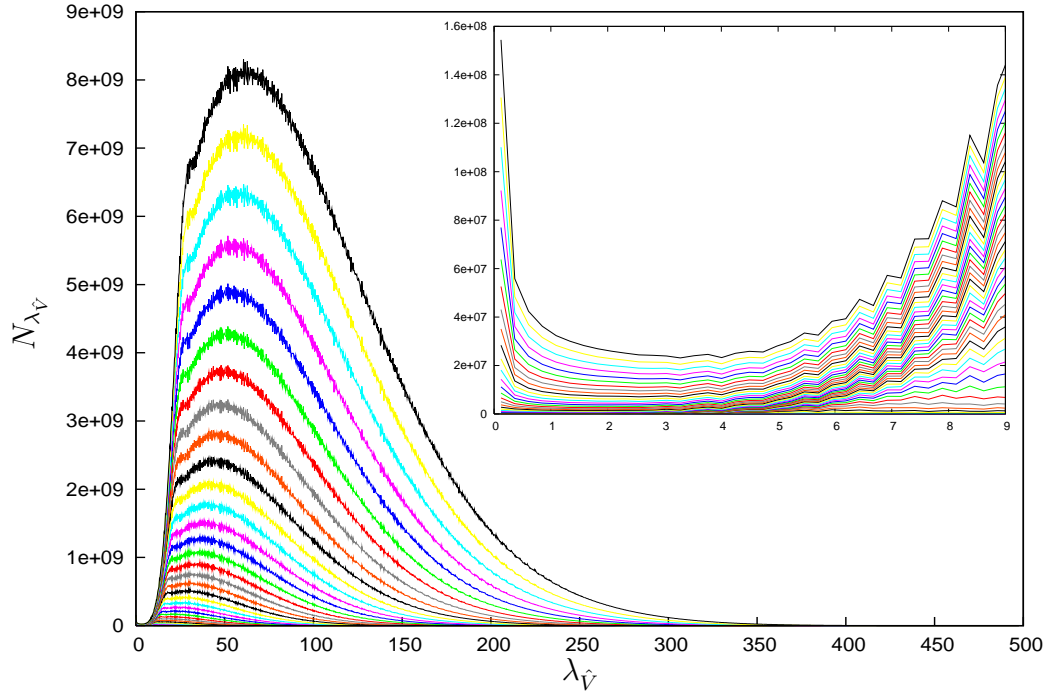}
\caption{\footnotesize(a) Overall 2048 bin histograms for the gauge invariant 5-vertex ($Z=\ell_P=1$) up to $j_{max}=\frac{44}{2}$(top curve, below it are histograms for
% $j_{max}=\frac{43}{2},\frac{42}{2},\ldots$)
smaller $j_{max}$).
There are $4.6 \times 10^{12}$ eigenvalues in all, of which $4.5\times 10^{11}$
are zero (and are excluded). % from the histogram).
 (b) Portion of histogram for $\eval\leq 9$.}
% 4680292624128 eigenvalues in all (including zeros)
% 450641502288 zeros in all
\label{pic}
\end{center}
\end{figure}

The action of $\hat{V}(R)$ on an arbitrary SNF decays into a sum over single vertices, so it is sufficient to compute its spectrum for a single vertex only.
We have implemented the matrices $\hat{q}_{IJK}$ for  a single $SU(2)$-gauge invariant $N_v$-valent vertex $v$ on a supercomputer. % in a $SU(2)$-gauge invariant setup. 
Here techniques from recoupling theory of angular momenta for the construction of a gauge invariant SNF as linear combinations of the $T_{\gamma\vec{j}\vec{m}\vec{n}}(\cdot)$ are heavily used: The gauge invariant subspace contained in the span of the SNF is computed by considering all ways to recouple the angular momenta of the edges incident at $v$ to a resulting trivial representation of $SU(2)$.
The second task is to examine which edge triple sign combinations $\vec{\epsilon}:=\{\epsilon(IJK)\}$ are realizable in an embedding of $N_v$ edges. There are
$\footnotesize N_v \choose 3$ triples $\epsilon(IJK)=0,\pm 1$ resulting in $3^{\mbox{\tiny $N_v \choose 3$}}$ possibilities. However for valences $>4$ not all of these possibilities can be realized. We have computed the set of realizable sign combinations $\vec{\epsilon}$
by a Monte Carlo 
random sprinkling of $N_v$ points on a unit sphere, where each point is
regarded as the end point of a vector emanating from the origin. The according $\epsilon(IJK)$-factors can then be computed. We exclude coplanar edge triples $\epsilon(IJK)=0$ from our analysis, as such configurations will never arise via sprinkling. For  a 5-vertex with 10 triples %$N_{Tr}=10$ 
we find that only 384 out of $2^{10}$ possibilities can be realized.
% This value can also be reobtained analytically.
For valences $N_v=4,5,6,7$ we have computed the eigenvalues $\lambda_{\hat{V}}$ for the %all possible 
matrices $\hat{V}$ for all sets of spins $j_1,\ldots,j_{N_v}\le j_{max}$ and 
all realizable
$\vec{\epsilon}$-sign configurations. Here $j_{max}$ is an upper cutoff. The $\lambda_{\hat{V}}$ can then be sorted into histograms to obtain a notion of spectral density. % depending e.g.\ on a fixed sign combination $\vec{\epsilon}$. 
We find that the spectral properties of $\hat{V}$ depend strongly on the $\vec{\epsilon}$. In particular one can choose $\vec{\epsilon}$ such that the smallest non-zero eigenvalues either 
increase, decrease or stay constant as $j_{max}$ is increased. There are also  $\vec{\epsilon}$-configurations for which all $\lambda_{\hat{V}}=0$ {\it independently} of the spins, as a consequence of gauge invariance. Figure \ref{pic} shows the resulting overall histogram for the gauge invariant $5$-vertex where all $\lambda_{\hat{V}}$ for all 384 $\vec{\epsilon}$ configurations are collected. For large eigenvalues ($>10$) we obtain a rapidly increasing eigenvalue density which can be fitted by an exponential.  For smaller eigenvalues ($\sim 3)$ the density becomes minimal and then increases again %without bound 
close to zero. This suggests that zero is an accumulation point of the volume spectrum. This property is shared by 6 and 7-valent vertices. The complete results can be found in a forthcoming paper \cite{NumVolSpec}.

% \section*{Acknowledgments}
\vspace{2mm}
\noindent \textbf{Acknowledgments}\hspace{3mm}
We thank Thomas Thiemann for encouraging discussions as well as the Numerical Relativity group of the Albert Einstein Institute Potsdam. J.B. thanks the Gottlieb Daimler- and Karl-Benz-foundation for financial support.  The work of D.R. was supported by the %Marie Curie Research and Training network 
European Network on Random Geometry, ENRAGE 
(MRTN-CT-2004-005616).

\end{document}